# Synchrobetatron resonance of crab crossing scheme with large crossing angle and finite bunch length

Derong Xu[*] and Yue Hao[†]
*Michigan State University, East Lansing, Michigan 48824, USA*

Yun Luo
*Brookhaven National Laboratory, Upton, New York 11973, USA*

Ji Qiang
*Lawrence Berkeley National Laboratory, Berkeley, California 94720, USA*

Crab crossing scheme is an essential collision scheme to achieve high luminosity for the future colliders with large crossing angles. However, when bunch length of one or both colliding beams is comparable with the wavelength of the crab cavity voltage, the nonlinear dependence of the crabbing kick may present a challenge to the beam dynamics of the colliding beams and impact the beam quality and the luminosity lifetime. In this paper, the results of nonlinear dynamics in the crab crossing scheme are presented, using both analytical and numerical approaches. The result indicates that higher-order synchrobetatron resonances may be excited in the crab crossing scheme with large crossing angle, which causes the beam quality deterioration and luminosity degradation. The studies also reveal possible countermeasures to suppress the synchrobeta resonance, hence mitigate the degradation of beam quality and luminosity.

DOI: 10.1103/PhysRevAccelBeams.24.041002

## I. INTRODUCTION

A crossing angle at the interaction region (IR) allows for a fast beam separation and smaller beta function at the interaction point (IP). The crossing angle leads to the geometric luminosity loss. A parameter to characterize the loss is the "Piwinski Angle" $\theta_P$:

$$\theta_P = \frac{\sigma_z}{\sigma_x} \theta_c \quad (1)$$

where $\sigma_{z/x}$ are the RMS longitudinal/horizontal bunch size and $\theta_c$ is the half crossing angle. For the symmetric-collider case with $\sigma_y \ll \sigma_x$, the geometric luminosity loss [1] is

$$R_L \equiv \frac{\mathcal{L}}{\mathcal{L}_0} = \sqrt{\frac{2}{\pi}} a e^b K_0(b),$$
$$a = \frac{\beta_y^*}{\sqrt{2}\sigma_z}, \qquad b = a^2(1+\theta_P^2) \quad (2)$$

where $\beta_y^*$ is the vertical beta function at IP and $K_0$ is the Bessel function.

To prevent the geometric loss, the crab cavities are employed to recover the luminosity [2]. The crab cavities introduce the transverse offset which depends on the longitudinal coordinate at IP. The dependence tilts the beam by half of the crossing angle to create an effective head-on collision. The crab crossing scheme can be achieved either by one crab cavity to excite a closed orbit as function of the longitudinal coordinate throughout the collision, or by a pair of crab cavities that constrain the tilting effect within the vicinity of IR.

The crab crossing scheme was first successfully implemented at the KEK B-factory [3], where a world record luminosity of $2.1 \times 10^{34}$ cm$^{-2}$ s$^{-1}$ was obtained in this configuration. The scheme is also planned in the LHC high luminosity upgrade [4], and as an essential design concept of the Electron Ion Collider, to achieve desired luminosity ($1 \times 10^{34}$ cm$^{-2}$ s$^{-1}$) [5].

Previous study [6] discussed the dynamics for different crab cavity frequency and the impact to the dynamic aperture for the LHC high luminosity upgrade. Our studies will focus on the dynamics for the electron-ion collision, which usually has much larger crossing angle and unsymmetric colliding beams.

In the IR region, as detailed in Sec. II A, each colliding beam will have a transverse offset as function of the location away from its reference particle $z$

[*]dxu@bnl.gov
Present address: Brookhaven National Laboratory, Upton, New York 11973, USA.
[†]haoy@frib.msu.edu

$$\Delta x(z) \sim \frac{\theta_c}{k_c} \sin(k_c z) - \theta_c z$$

where $k_c = 2\pi f_c/c$ is the wave number of the crab cavity. Ideally when the frequency of the crab cavity is low enough or the bunch is very short ($k_c \sigma_z \ll 1$), the transverse deviation, scaled as $k_c^2 z^3$, is sufficiently small and the crab kick fully compensates the effect of the crossing angle. However, the cavity with very low frequency has large surface area and is not feasible to be manufactured and processed with current technology of super-conduction rf cavity. The bunch length of the beam, especially the ion beam, is usually not short enough, due to longitudinal dynamics and lack of an effective cooling method. Therefore it is essential to study the beam dynamics consequence when $k_c \sigma_z \sim 1$. We refer to this as the finite bunch length effect, and will study the dynamics of such effect using an example with the parameters similar to that of the future Election-Ion Collider (EIC) which adopts the crab crossing scheme. The bunch length of the ion beam is usually longer than that of the electron beam. Meanwhile, the damping of the ion beam (via hadron cooling) is much weaker than the synchrotron radiation damping of the electron beam. The cooling time of the hadron beam is expected to be 1 hour even in the presence of strong hadron cooling [7], compared with < 100 millisecond damping time for the electron beam. Therefore, we will focus on the effect of the ion beam in this article.

The dynamics of the crossing scheme are analyzed with various approaches in this article, including the nonlinear driving terms of one colliding beam with thin lens approximation of beam-beam interaction, the strong-strong self-consistent simulations and the frequency map analysis using the turn-by-turn tracking data. The content is structured as follows. In Sec. II we first introduce the bunch length effect in the presence of crab cavities and crossing angle. Then the strength of synchrobetatron resonance driving terms of the ion beam is derived with a thin and strong model of the electron beam. In Sec. III we present the simulation results with weak-strong and strong-strong code and apply frequency map analysis to identify the involved synchrobetatron resonances. In Sec. IV, three possible methods are brought up to reduce the luminosity degradation which are also supported by simulation. The conclusion is given in Sec. V.

## II. THEORETICAL MODEL

In this section, the model of the beam-beam analysis is reviewed. They are used to retrieve the nonlinear driving terms of the beam-beam interaction when the opposing beam is assumed to be a thin length element.

### A. Crossing angle

In the laboratory frame, the $s$ axes of the two beams are along their moving directions. When there is a crossing angle in the collision, the beam-beam field generated by the other beam depends not only on the transverse position but also on the longitudinal coordinate. It is convenient to transform the laboratory frame to a head-on frame by a Lorentz boost [8]

$$\tilde{p}_x = \frac{p_x - h \tan \theta_c}{\cos \theta_c}, \qquad \tilde{p}_y = \frac{p_y}{\cos \theta_c}$$
$$\tilde{p}_z = p_z - p_x \tan \theta_c + h \tan^2 \theta_c,$$
$$\tilde{x} = x\left[1 + \left(\frac{\partial \tilde{h}}{\partial \tilde{p}_x}\right) \sin \theta_c\right] + z \tan \theta_c$$
$$\tilde{y} = y + x\left(\frac{\partial \tilde{h}}{\partial \tilde{p}_y}\right) \sin \theta_c,$$
$$\tilde{z} = \frac{z}{\cos \theta_c} + x\left(\frac{\partial \tilde{h}}{\partial \tilde{p}_z}\right) \sin \theta_c, \qquad (3)$$

where

$$h = 1 + p_z - \sqrt{(1+p_z)^2 - p_x^2 - p_y^2}, \qquad (4)$$

and a variable marked with a superscript tilde means quantity in the head-on frame. Here $x$, $y$ and $z = -c\Delta t$ are transverse coordinates and arriving time relative to the reference particle respectively, $p_{x,y}$ are the transverse momentum normalized by the momentum $p_0$ of the reference particle, and $p_z$ is the energy deviation normalized by $p_0 c$.

The linear part can be written as a Lie operator

$$\mathcal{M}_L = e^{\lambda} e^{-:z p_x \tan \theta_c:} e^{:\lambda(x p_x + y p_y - z p_z):} \qquad (5)$$

where $\lambda$ is a scale coefficient

$$e^{\lambda} = \frac{1}{\sqrt{\cos \theta_c}} \qquad (6)$$

To provide a head-on collision at the IP, a thin crab cavity is used to tilt the colliding bunch. The thin crab cavity is placed with $\pi/2$ phase in the horizontal plane and thus imparts a transverse momentum kick

$$\Delta p_x = -\frac{\tan \theta_c}{k_c \sqrt{\beta_{x,cc} \beta_x^*}} \sin(k_c z) \qquad (7)$$

where $\beta_{x,cc}$, $\beta_x^*$ are horizontal beta functions at the crab cavity and IP. The symplecticity of the crab kick demands that the off axis particle will receive an energy kick by the crab cavity,

$$\Delta p_z = -\frac{x \tan \theta_c}{\sqrt{\beta_{x,cc} \beta_x^*}} \cos(k_c z) \qquad (8)$$

Equations (7)–(8) can also be described by Lie notation

$$\mathcal{M}_{cc} = \exp\left[-:\frac{x\tan\theta_c}{k_c\sqrt{\beta_{x,cc}\beta_x^*}}\sin(k_cz):\right] \quad (9)$$

After collision, an identical crab cavity is used to restore the tilt. The Lie maps before and after collision are

$$\mathcal{M}_b = \mathcal{B}\left(-\frac{\pi}{2}\right)\mathcal{M}_{cc}\mathcal{B}\left(\frac{\pi}{2}\right)\mathcal{M}_L$$

$$= e^{\lambda}e^{:p_x\tan\theta_c[\frac{\sin(k_cz)}{k_c}-z]:}e^{:\lambda(xp_x+yp_y-zp_z):}$$

$$\mathcal{M}_a = \mathcal{M}_L^{-1}\mathcal{B}\left(\frac{\pi}{2}\right)\mathcal{M}_{cc}\mathcal{B}\left(-\frac{\pi}{2}\right)$$

$$= e^{-\lambda}e^{-:\lambda(xp_x+yp_y-zp_z):}e^{-:p_x\tan\theta_c[\frac{\sin(k_cz)}{k_c}-z]:}$$

$$= \mathcal{M}_b^{-1} \quad (10)$$

where

$$\mathcal{B}\left(\pm\frac{\pi}{2}\right) = \exp\left[\mp:\frac{\pi}{2}\left(\frac{x^2}{\sqrt{\beta_{x,cc}\beta_x^*}}+p_x^2\sqrt{\beta_{x,cc}\beta_x^*}\right):\right]$$

are betatron linear maps from IP (crab cavity) to crab cavity (IP).

In the head-on frame, the horizontal coordinate at the IP is

$$x_1 = \mathcal{M}_b x \approx x - \tan\theta_c\left[\frac{\sin(k_cz)}{k_c}-z\right] \quad (11)$$

which indicates that a particle longitudinally deviating from the bunch center will have a horizontal offset. Figure 1 presents the beam distribution at the IP in normalized $x$–$z$ plane. The distribution is determined by two dimensionless parameters: $k_c\sigma_z$ and Piwinski angle $\theta_P$. To minimize the tilting effect, smaller $k_c\sigma_z$ and $\theta_P$ are more desirable.

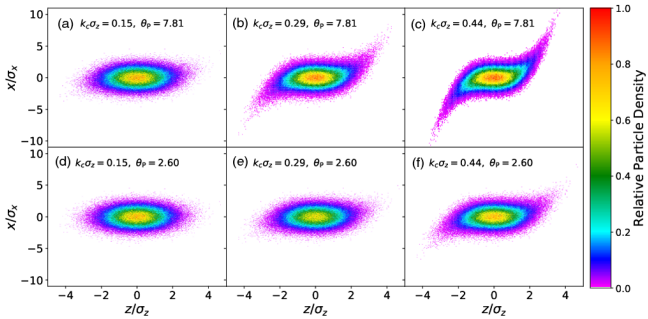

FIG. 1. The horizontal tilting effect at IP for different parameters: $k_c\sigma_z$ and $\theta_P$. $k_c$ is the wave number of the crab cavity, $\sigma_z$ the bunch length, and $\theta_P$ is the Piwinski angle.

### B. Beam-beam interaction

We use a weak-strong model to simplify our discussion. In this model, the electron beam is treated as a strong and thin beam that collides with the reference ion particle at the IP. This is a good approximation when electron distribution changes very slowly after several damping time.

The following discussion is within the head-on frame enabled by the Lorentz boost, and we drop all tilde sign for all variables for simplicity.

A particle with longitudinal coordinate $z$ in the ion beam collides with the electron beam at $s = z/2$. The collision process can be described by Hirata's synchrobeam map [9]

$$\mathcal{M}_{bb} = \mathcal{D}\left(\frac{z}{2}\right)\exp(-:U:)\mathcal{D}\left(-\frac{z}{2}\right) \quad (12)$$

where

$$\mathcal{D}\left(\pm\frac{z}{2}\right) = \exp\left[\mp:\frac{z}{4}(p_x^2+p_y^2):\right] \quad (13)$$

are drift operators. Assuming that the strong beam has an upright bi-Gaussian distribution, the beam-beam potential produced by the strong beam is

$$U = \frac{Q_1Q_2Nr_0}{\gamma_0}\int_0^\infty du\,\frac{\exp\left(-\frac{x^2}{2\sigma_x^2+u}-\frac{y^2}{2\sigma_y^2+u}\right)}{\sqrt{2\sigma_x^2+u}\sqrt{2\sigma_y^2+u}} \quad (14)$$

Here $N$ is the total particle number of the strong beam, $r_0 = e^2/(4\pi\epsilon_0mc^2)$ is the classical radius, $\gamma_0$ is the relativistic factor of the weak beam, $Q_{1,2}$ are the charge numbers of particles from two colliding bunches, and $\sigma_{x,y}$ are the RMS beam sizes of the strong beam at the collision point.

Including the Lorentz boost and the crab cavity, the total map for beam-beam interaction is

$$\mathcal{M}_{\mathrm{IP}} = \mathcal{M}_b\mathcal{M}_{bb}\mathcal{M}_a = \exp(-:\mathcal{TSU}:) \quad (15)$$

in which the scaling operator $\mathcal{S}$ is

$$\mathcal{S} = \exp[:\lambda(xp_x+yp_y-zp_z):] \quad (16)$$

and the transformation operator $\mathcal{T}$ is

$$\mathcal{T} = \exp\left[:p_x\left(\frac{\sin(k_cz)}{k_c}-z\right)\tan\theta_c - \frac{e^{3\lambda}}{4}z(p_x^2+p_y^2):\right] \quad (17)$$

The relation $\theta_c \ll 1$ is usually fulfilled which leads to

$$\mathcal{S} \approx 1, \quad e^{3\lambda} \approx 1$$

If we denote $f(z)$ as

$$f(z) = -\tan\theta_c \left[\frac{\sin(k_c z)}{k_c} - z\right] \quad (18)$$

we have

$$\mathcal{TS}z = z = z_{CP}, \qquad \mathcal{TS}y = y + \frac{1}{2}zp_y = y_{CP},$$

$$\mathcal{TS}x = x + \frac{1}{2}zp_x + f(z) = x_{CP} + f(z_{CP})$$

where the subscript "CP" indicates that the variables are evaluated at the collision point. As a result,

$$U_{bb} = \mathcal{TS}U(x, y; \sigma_x, \sigma_y)$$
$$= U(x_{CP} + f(z_{CP}), y_{CP}; \sigma_{x,CP}, \sigma_{y,CP}) \quad (19)$$

The Hamiltonian of a test particle in the weak beam is

$$H = \frac{1}{2}(p_x^2 + K_x x^2) + \frac{1}{2}(p_y^2 + K_y y^2)$$
$$+ U(x + f(z), y; \sigma_x, \sigma_y)\delta\left(s - \frac{z}{2}\right) \quad (20)$$

The lattice focusing structures are described by $K_x(s)$ and $K_y(s)$. The $\delta$-function represents the periodic collisions modulated by the longitudinal motion.

### C. Synchrobetatron resonances

As the astute reader may already notice, Eq. (20) did not include the dynamics of longitudinal and redeem $z$ as a parameter. It reflects our intent to simplify the question to a 4D case by treating longitudinal coordinate $z$ as a harmonic oscillator. It is a sensible approximation since the RMS longitudinal action of a beam is much larger than the corresponding RMS actions of the transverse planes [10] and, in consequence, the beam-beam interaction has a negligible impact on the longitudinal motion.

The beam-beam potential is expanded in Taylor series in order to study the synchrobetatron resonances

$$U(x + f(z), y; \sigma_x, \sigma_y) = a_{00} + a_{10}x + a_{20}x^2 + a_{02}y^2$$
$$+ a_{30}x^3 + a_{12}xy^2 + a_{40}x^4$$
$$+ a_{22}x^2y^2 + \cdots \quad (21)$$

Here the coefficients $a_{mn}$, as well as parameters $\sigma_{x,y}$, are functions of $z$.

The zero order $a_{00}$ is

$$a_{00} = \frac{Q_1 Q_2 N r_0}{\gamma_0} \int_0^\infty du \frac{\exp\left(-\frac{f^2}{2\sigma_x^2 + u}\right)}{\sqrt{2\sigma_x^2 + u}\sqrt{2\sigma_y^2 + u}} \quad (22)$$

where $f(z)$ in Eq. (18) is abbreviated as $f$ without confusion. Note here that $a_{00}$ diverges because we define potential at infinity as zero. It has no effect on transverse motions.

The first order corresponds to a dipole kick, which distorts the horizontal closed orbit. The orbit distortion is as small as submicron range [11], and its dynamic effect is omitted in this paper. The second orders give rise to linear beam-beam tune shift. All higher orders as well as the first order drive synchrobetatron resonances. Appendix shows how to calculate $a_{mn}$ to a very high order.

For a real beam, particles form a finite distribution around the bunch center, i.e.,

$$|x| < 5\sigma_x, \quad \text{and} \quad |y| < 5\sigma_y$$

Therefore, the series of the potential, Eq. (21), can be truncated at a specific order. The truncated technique has been used in the study of beam-beam interaction for round beams [12]. Figure 2, which presents beam-beam kicks for particles at different positions, demonstrates that it is also useful for nonround beams. Due to the nature of the potential for particle interactions and the additional deviation due the to crab crossing scheme, the truncation has to be done beyond 60th order. With the help of increased numerical capability, we are able to truncate the series at order of 100 or more. We then rewrite Eq. (21) as

$$U(x + f(z), y; \sigma_x, \sigma_y) = \sum_{m=0}^{M}\sum_{n=0}^{N} a_{mn} x^m y^n \quad (23)$$

After Floquet transformation, the new Hamiltonian can be expressed in terms of action angle variables in the form

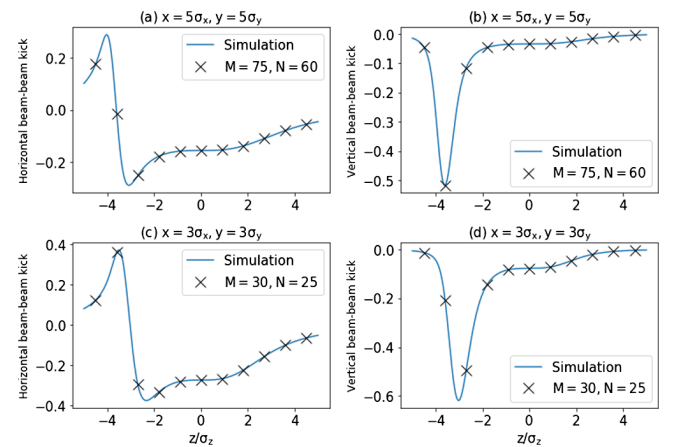

FIG. 2. Beam-beam kick for particles with different horizontal coordinates. The blue curves are calculated by tracking while the discrete points are calculated by a truncated power series. $\sigma_x$, $\sigma_y$, and $\sigma_z$ are beam sizes. $M$ and $N$ are truncated orders in Eq. (23). All vertical axes are normalized by the same value.

$$H = \nu_x J_x + \nu_y J_y + \frac{1}{2} h_{00}(J_x, J_y; z)$$
$$+ \sum_{m=0}^{M} \sum_{n=-N}^{N} \sum_{l=-\infty}^{\infty} h_{mn}(J_x, J_y; z) \cos(m\psi_x + n\psi_y + l\theta) \quad (24)$$

here $\theta$ is the orbit angle in the accelerator and $h_{mn}$ can be obtained from $a_{mn}$ as,

$$h_{mn} = 2 \sum_{i=0}^{\lfloor \frac{M-m}{2} \rfloor} \sum_{j=0}^{\lfloor \frac{N-|n|}{2} \rfloor} a_{m+2i,|n|+2j} C_{m+2i}^{i} C_{|n|+2j}^{j}$$
$$\times \left(\frac{1}{2}\beta_x J_x\right)^{i+\frac{m}{2}} \left(\frac{1}{2}\beta_y J_y\right)^{j+\frac{|n|}{2}} \quad (25)$$

where the symbol C denotes the binomial coefficient and $\lfloor \cdots \rfloor$ represents the rounding down operation to nearest integer.

As $z$ is a harmonic oscillator with tune $\nu_z$, the coefficients $h_{mn}$ in Eq. (24) can be expanded in Fourier series:

$$h_{mn} = \sum_{p=-\infty}^{\infty} c_{mnp} \cos(p\nu_z \theta) + s_{mnp} \sin(p\nu_z \theta)$$
$$= \sum_{p=-\infty}^{\infty} \hat{h}_{mnp} \cos(p\nu_z \theta + \chi_{mnp}) \quad (26)$$

where

$$c_{mnp} = \frac{1}{2\pi} \oint d\theta h_{mn} \cos(p\nu_z \theta)$$
$$s_{mnp} = \frac{1}{2\pi} \oint d\theta h_{mn} \sin(p\nu_z \theta)$$

Substituting Eq. (26) into Eq. (24), the Hamiltonian becomes

$$H = \nu_x J_x + \nu_y J_y + \frac{1}{2} \sum_{p=-\infty}^{\infty} \hat{h}_{00p}(J_x, J_y) \cos(p\nu_z \theta + \chi_{00p})$$
$$+ \sum_{m=0}^{M} \sum_{n=-N}^{N} \sum_{p=-\infty}^{\infty} \sum_{l=-\infty}^{\infty} \hat{h}_{mnp}(J_x, J_y)$$
$$\times \cos(m\psi_x + n\psi_y + p\nu_z \theta + l\theta + \chi_{mnp}) \quad (27)$$

A single betatron resonance therefore splits into a spectrum of resonances, and the resonance conditions are

$$m\nu_x + n\nu_y + p\nu_z + l = 0 \quad (28)$$

The corresponding resonance strength is $\hat{h}_{mnp}$. Because the beam-beam potential is an even function about $y$, the conditions

$$a_{mn} = 0, \quad h_{mn} = 0$$

hold when $n$ is odd. Resonances with odd $n$ in Eq. (28) are not excited. However, these conditions no longer hold with $x$ because of the horizontal tilting effect in Fig. 1. Resonances with odd $m$ are excited.

Therefore, due to the finite bunch length effect, a new set of the resonance with odd $m$ will be excited. In the next section, we will concentrate on these new set of resonances.

### III. BEAM-BEAM SIMULATION RESULTS

Although the analysis of the synchrobetatron resonance in the previous section provides insights on the physics of crab-crossing beam-beam interaction, the multiparticle simulation is an irreplaceable tool to study the beam quality degradation of both colliding beams and its consequence of the luminosity. The parameters used in simulation are shown in Table I. The parameters have the following features: (i) Beam-beam parameter for the ion beam is less than 0.015 (ii) Beam-beam parameter for the electron beam is less than 0.1 (iii) Large crossing angle, 25 mrad between ion and electron trajectory (iv) Both beams are flat at IP for the beam size matching of both colliding beams, with flatness $r = \sigma_y/\sigma_x \sim 0.2$ (v) Small vertical beta function for both beams. (vi) Ion beam are longer than the electron beam.

We use strong-strong (SS) simulation code BeamBeam 3D [13] and its hybrid weak-strong (HWS) variation to study the crab crossing beam dynamics caused by the finite bunch length effect. The SS simulation uses a self consistent approach so that it allows both beam distributions affected by beam-beam interaction. In BB3D, a particle-in-cell (PIC) Poisson solver, which uses a computational grid to obtain the charge density distribution, is used to calculate the beam-beam force from an arbitrary beam distribution. The SS simulation was used widely to predict if there is a coherent beam-beam mode. With proper tune choice of both beams, coherent instability can be avoided in SS simulation. However, SS simulation is still useful in the first several damping time, to model the evolution of the electron distribution and its impact on the proton beam before the electron beam reaches equilibrium. SS simulation is also known for its higher numerical noise. To minimize the noise effect, a special HWS version of BeamBeam3D is developed to start with SS simulation and memorize the electron's distribution at $N$th turn and reuse the distribution information in latter turns. The HWS is expected to provide a better prediction of the long term evolution of ion beam's quality.

TABLE I. Related parameters of Electron-Ion Collider in this paper. "H" stands for horizontal and "V" denotes vertical below. Synchrotron radiation and damping are not considered for protons.

| Parameter | Proton | Electron |
|---|---|---|
| Circumference [m] | 3833.8 | |
| Energy [GeV] | 275 | 10 |
| Particles per bunch [$10^{11}$] | 1.04 | 3.44 |
| Crossing angle [mrad] | 25.0 | |
| Crab cavity frequency [MHz] | 200.0 | 400.0 |
| $\beta_x^*/\beta_y^*$ [cm] | 90.0/5.90 | 72.0/10.2 |
| RMS emittance (H/V) [nm] | 16.0/8.50 | 20.0/4.92 |
| RMS bunch size (H/V) [$\mu$m] | 120.0/22.4 | |
| RMS bunch length [cm] | 7.0 | 2.0 |
| RMS energy spread [$10^{-4}$] | 6.6 | 5.8 |
| Transverse fractional tune (H/V) | 0.310/0.305 | 0.08/0.06 |
| Synchrotron tune | 0.010 | 0.069 |
| Transverse damping time [turns] | $\infty$ | 4000 |
| Longitudinal damping time [turns] | $\infty$ | 2000 |
| Beam-beam parameter (H/V) | 0.015/0.005 | 0.10/0.08 |
| Parameter $k_c \sigma_z/\theta_P$ | 0.29/7.29 | 0.17/2.08 |

The number of macroparticles used in the simulations are 0.5 and 2 million for electron and proton bunch respectively. The electron bunch is split into 7 slices longitudinally while the proton bunch is cut into 25 slices. These numbers are selected by converging studies with the comprise of the numerical noise and the simulation time. Other parameters, such as the computational grid number in BB3D, are determined empirically.

In the following simulation studies, the comparison between different beam parameters are used to unveil the dynamics of finite bunch length effect in crab crossing scheme, with the same simulation scheme and parameters. Although the choice of simulation parameter may affect the computing time or the precision of results, it does not change the physical nature.

### A. Degradation

Figure 3 presents the comparison between crab crossing and head-on collision with the same parameters except for the crossing angle and crab cavity amplitude using the SS simulation. The luminosity shows a similar trend in both collision schemes. A fast initial drop happens at the first few thousand turns, and then the degradation converges to a finite negative slope. The initial quick drop of the luminosity is mostly contributed by the dynamic beta effect of the electron beam, especially in the vertical plane. After the initial drop, a quicker trend of luminosity degradation is observed in the crab crossing collision, compared with that in the head-on case.

The degradation rate in Fig. 3 as well as in all other luminosity evolution figures in this paper, is fitted by the last 60% luminosity data. Assuming the fitting is described by

$$\mathcal{L} = \mathcal{L}_0 + KN \quad (29)$$

where $\mathcal{L}$ is the tracked luminosity while $N$ is the tracked turn. Then the degradation rate is defined as

$$\mathrm{DR} \equiv \frac{K}{\mathcal{L}_0} \quad (30)$$

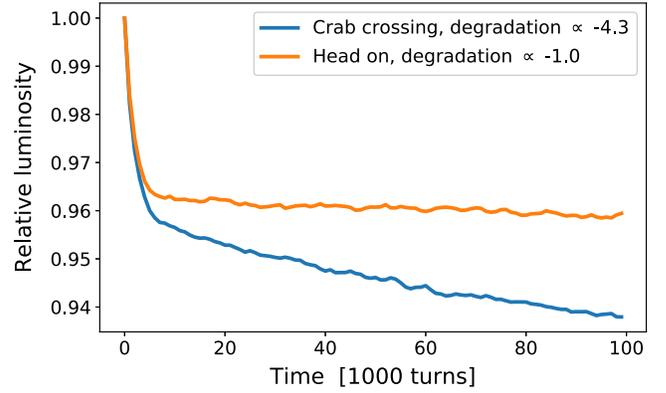

FIG. 3. The luminosity evolution as function of time by SS simulation for crab-crossing and head-on cases, the luminosity data is normalized by its average value of the first 1000 turns in the simulation while the degradation rates are normalized by the head-on SS number.

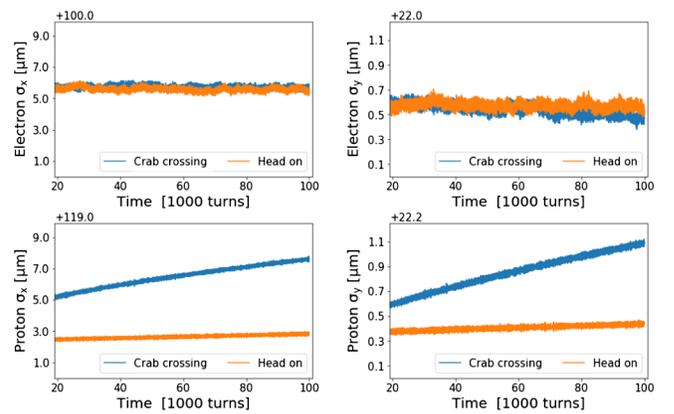

FIG. 4. The beam size evolution with crab crossing scheme from SS simulation.

The degradation rate is also normalized by the head-on SS simulation number to make DR comparable across different figures.

As illustrated in Fig. 3, SS simulation predicts higher degradation rate in the crab-crossing scheme than that in the head-on scheme. This motivates the following studies on the root cause of such a difference. Figure 4 well illustrates that the electron beam reaches its equilibrium after several damping time (around 20K revolutions) and only ion beam size changes visibly. The equilibrium electron beam sizes change slowly since the beam-beam strength from the ion beam is altered due to the ion beam size changes.

The numerical noise effect in the SS beam simulation is known to cause larger artificial emittance growth than the weak strong counterpart. The same calculation is done using the new HWS version of Beambeam3D. The weak strong simulation starts after 60K turns of SS simulation and continues for 40K turns. Figure 5 and Figure 6 show the evolution of luminosity and beam sizes of the total 100K simulation. The trend from linear fit only uses the last 40K weak-strong portion.

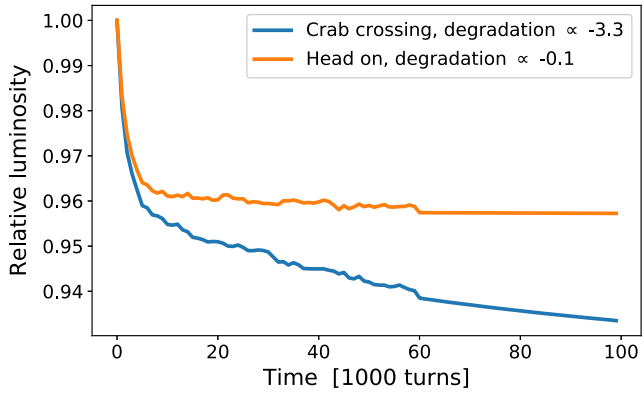

FIG. 5. The luminosity evolution as function of time by HWS simulation for crab-crossing and head-on cases, the luminosity data is normalized by its average value of the first 1000 turns in the simulation while the degradation rates are normalized by the head-on SS number. The linear fitting uses the last 40000 turns weak-strong simulation data only.

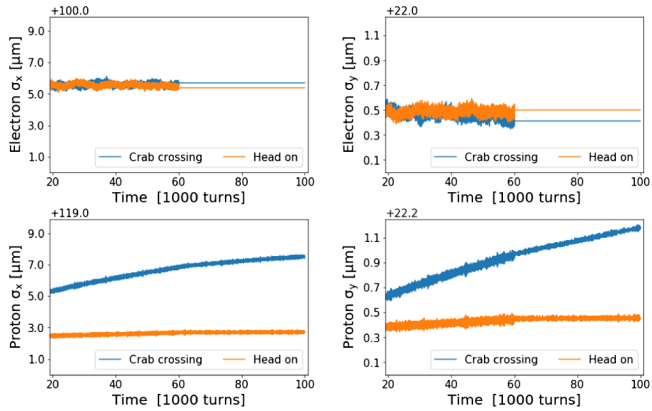

FIG. 6. The beam size evolution with crab crossing scheme from HWS simulation.

Although the emittance growth and luminosity degradation rates decreases when switching from SS to WS simulation, the drastic difference between the crab-crossing and the head-on remains. Therefore, in both SS and WS simulations, the proton beam emittance grows due to beam-beam interaction and causes the luminosity degradation, due to the lack of radiation damping.

### B. Scaling

The characteristics of the crab-crossing induced beam quality degradation can be further understood by the scaling study with beam-beam parameters and the beam parameters related to the finite bunch length effect of the ion beam (proton in this example).

A smaller beam-beam parameter of ion beam is expected to reduce the luminosity degradation rate. Starting from the parameters in Table I as a baseline, the beam-beam parameters of the ion beam vary with the population of the opposing bunch. The upper plot in Fig. 7 clearly verifies this dependence. However, the reduction of the beam-beam parameter of the electron beam does not clearly improve the degradation, as shown in the bottom plot in Fig. 7. It only results in significant difference of the initial luminosity drop within the damping time (4000 turns). The difference can be further understood by the beam size evolution of the electron beam, shown in 8. The electron equilibrium beam sizes are not changed when the beam-beam parameter of the ion beam is modified by the electron beam intensity, as indicated in the top figure. However, the equilibrium beam sizes, especially the vertical beam size are affected by the electron beam-beam parameter. The vertical equilibrium beam size shrinks as the beam-beam parameter decreases, which results in a smaller initial drop of the luminosity.

The root cause of this degradation can be demonstrated by the scaling study of the bunch length of the ion/proton beam and its crab cavity's frequency. Figure 9 illustrates that the luminosity degradation is reduced with shorter bunch length and smaller frequency of the crab cavity.

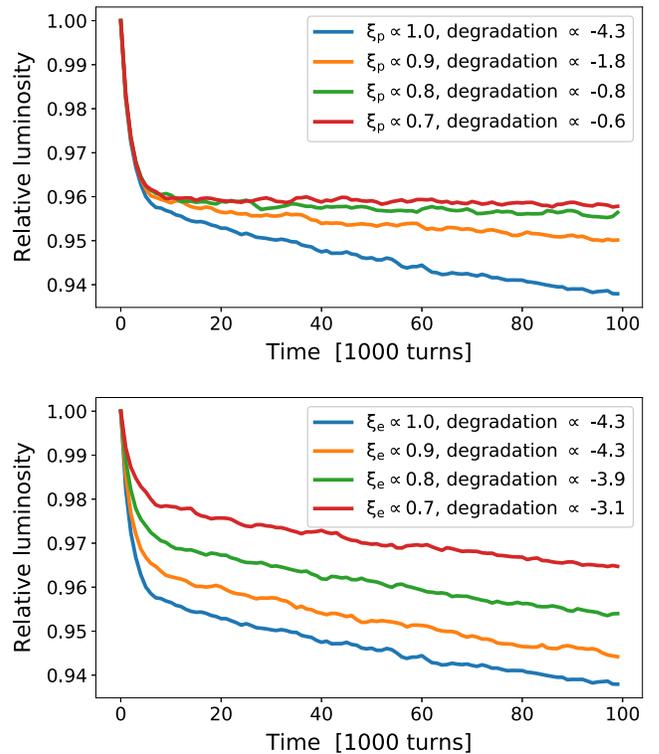

FIG. 7. Luminosity degradation with different beam-beam parameter of the proton beam (top figure) and of electron beam (bottom figure). The proton/electron beam-beam parameter $\xi_{p/e}$ is scaled by varying the opposite beam intensity.

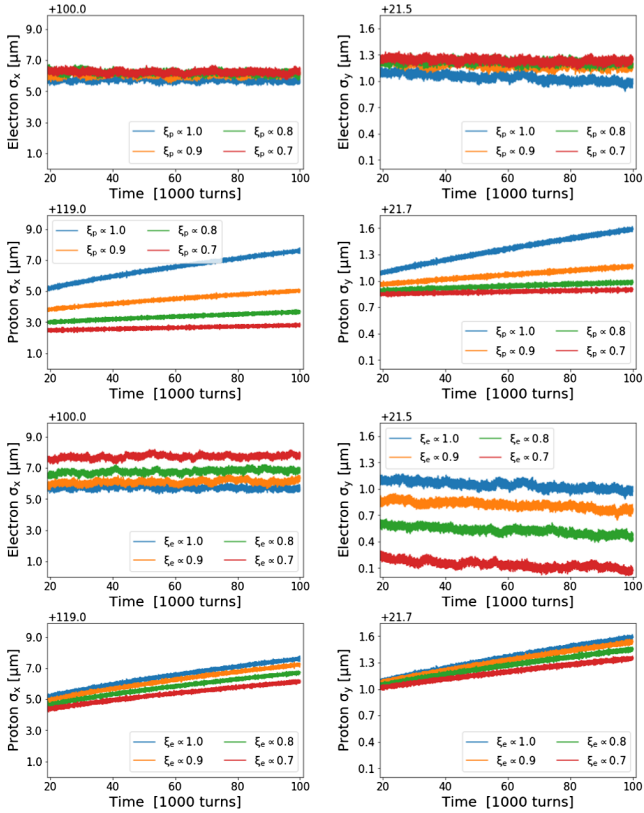

FIG. 8. Beam size evolution of different beam-beam parameter of the proton beam (top figure) and of electron beam (bottom figure). The proton/electron beam-beam parameter $\xi_{p/e}$ is scaled by varying the opposite beam intensity.

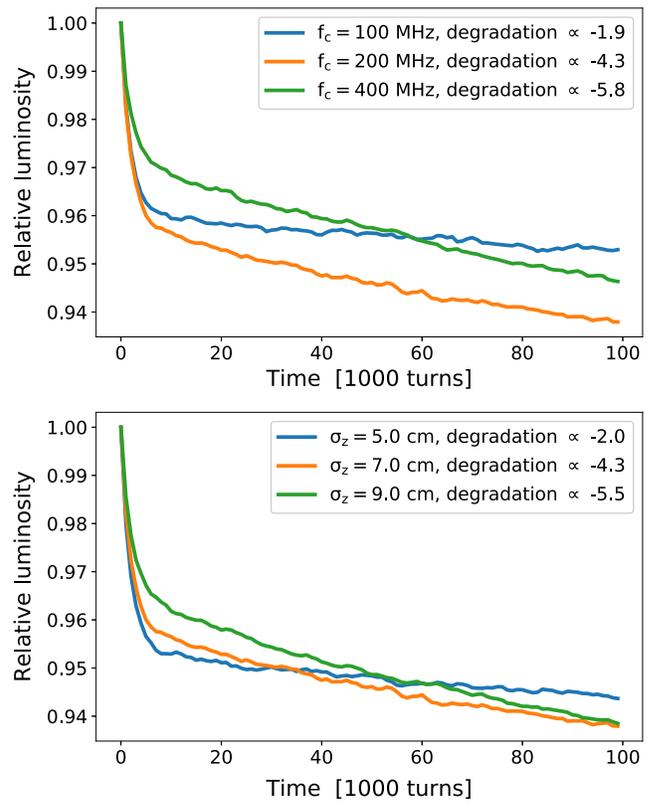

FIG. 9. Luminosity evolution with different crab cavity frequency (top figure) and bunch length (bottom figure) of the proton beam.

Recalling that for small $k_c \sigma_z$, the horizontal deviation $\Delta x(\sigma_z)$ caused by the crab cavity of ion beam is proportional to $k_c^2 \sigma_z^3$. Figure 10 indicates that starting from nominal values in Table I, the ion beam's emittance growth, as a function of the ion bunch length and the frequency, can be approximated as:

$$\frac{d\epsilon_{i,x}}{dt} = g(\sigma_{i,z}, f_c) \sim g(\sigma_{i,z}^3 f_c^2) \qquad (31)$$

Here $g$ denotes an unknown function. When $k_c \sigma_z$ becomes large enough (beyond the example parameter in Table I), the scaling behavior deviates from the above approximation.

It is clear that the transverse dynamics was strongly affected by the synchrotron motion in the crab crossing scheme. The scheme is similar to other studies about synchrobetatron resonance [14,15]. In synchrobetatron resonances, a smaller synchrotron tune is expected to make the resonance higher order, therefore reducing the effect of the resonance. The scaling of luminosity degradation with synchrotron tune in the crab-crossing scheme is illustrated in the Fig. 11. The luminosity degradation vanishes with an order of magnitude lower synchrotron tune, while is significant improved even at half synchrotron tune.

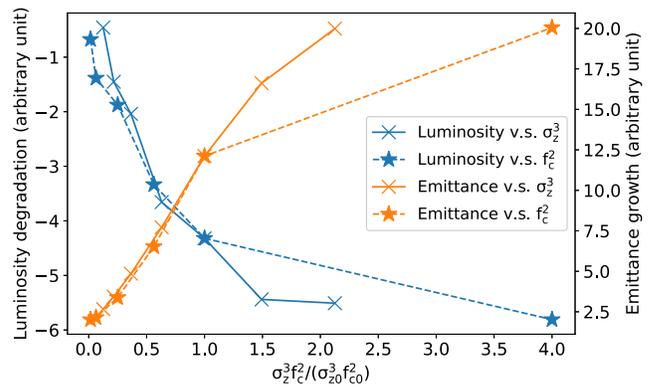

FIG. 10. The scaling relation of the luminosity degradation and emittance growth as functions of $\sigma_z^3 f_c^2$. Here $\sigma_{z0}$ and $f_{c0}$ are the nominal parameters in Table I.

### C. Frequency map analysis

FMA is a widely used method to explore dynamics of Hamiltonian systems [16]. It also turns out to be very useful in studying beam-beam effects [17,18]. In this paper FMA of the transverse motion is used in presence of the synchrotron oscillation, in order to find out the root cause of the luminosity degradation. It also helps to understand the scaling behavior of the degradation with the beam-beam parameter, finite bunch length and synchrotron tune of the ion beam.

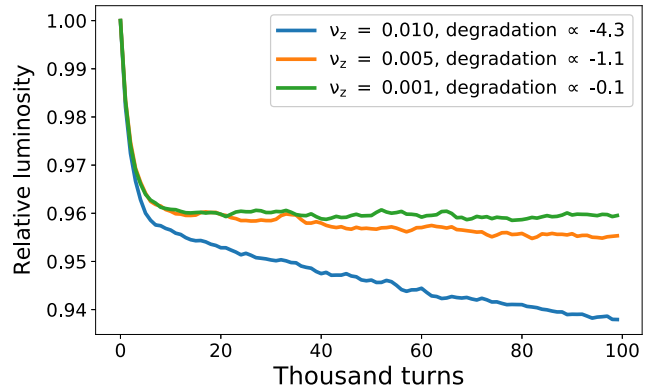

FIG. 11. Luminosity evolution for different longitudinal tune of proton beam.

In SS simulation, two colliding beams are tracked for 90,000 turns which is sufficiently long enough that two beams reach quasiequilibrium state in the presence of beam-beam interaction. Then, the turn-by-turn coordinates are collected for more than 60,000 proton macroparticles in next 1024 consecutive turns. Frequency is computed by NAFF algorithm [16] using the Hanning window of 500 turns (5 longitudinal period). Then the window is shifted 50 times by 10 turns. Particle tune takes the average of these frequencies with an interval of 100 turns (1 longitudinal period) and the diffusion index is calculated as

$$D = \log_{10}\sqrt{\sigma_{\nu_x}^2 + \sigma_{\nu_y}^2} \qquad (32)$$

where $\sigma_{\nu_x}$ and $\sigma_{\nu_y}$ are the RMS spread of those horizontal and vertical tunes.

An example of tune variation $\log_{10}(\sigma_\nu)$ of 1000 particles is shown in Fig. 12. The variation of longitudinal tune is far smaller than the transverse ones which is a direct proof that the longitudinal motion is barely affected by the transverse dynamics caused by the beam-beam interaction.

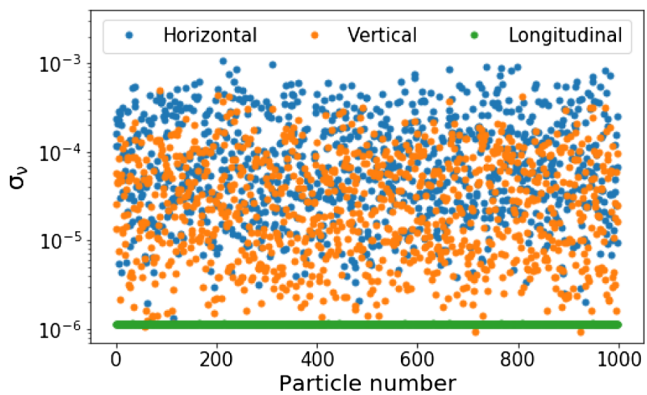

FIG. 12. The tune variation of 1000 particles in SS simulation.

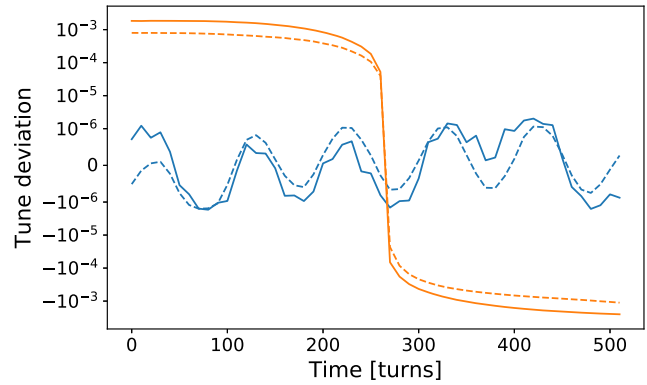

FIG. 13. Tune deviation from the average versus the window shift (in turns) for particles with different diffusion index: small (blue) and large (yellow). The solid lines give horizontal tune while the dashed lines show the vertical.

The frequencies evolve gradually as the window is shifted as shown in Fig. 13, where the horizontal axis represents the window shift in turns and the vertical one— the transverse tune deviation from the average in logarithmic scale. The blue curve, which oscillates in a sinusoidal pattern approximately, shows the deviation for the particle with a small diffusion index. The yellow curve represents the particle with a large diffusion index.

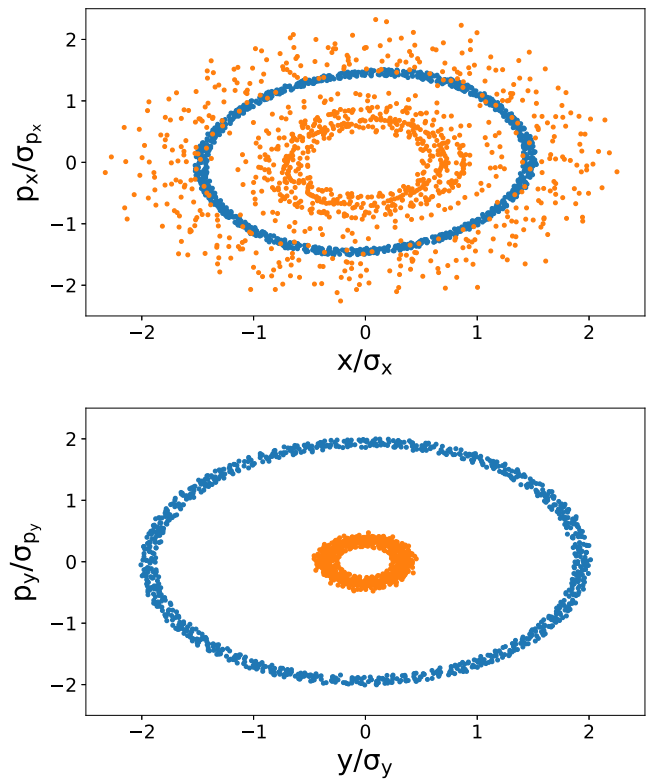

FIG. 14. Phase space trajectories for particles with different diffusion index: small (blue) and large (yellow). Upper and lower pictures are for horizontal and vertical plane, respectively.

The corresponding phase space trajectories for the two particles are plotted in Fig. 14. The blue trajectory remains elliptical within our tracking period. On the contrary, the yellow curve tends to less bounded and behaves similar to a chaotic motion. Although we use the term "chaotic" here, there is no particle loss during our tracking. All what we have observed is the rms emittance growth.

Figure 15 presents the frequency map with parameters in Table I by SS simulation. There are at least 2 kinds of synchrobetatron resonances existing in the vicinity of the working point (0.31,0.305)

$$3\nu_x + p\nu_z = 1$$
$$2\nu_x - 2\nu_y + p\nu_z = 0 \qquad (33)$$

as labeled in Fig. 15. The first kind of resonance only exists in crab crossing scheme. It is caused by the tilting effect as described in Sec. II. The second kind may arise due to the hourglass effect, especially in the vertical plane due to the smaller vertical beta function at the IP. The two resonances are excited by $h_{30}$ and $h_{2,-2}$ in Eq. (24) respectively.

Figure 16 shows the distribution of more than 60,000 protons in longitudinal phase space for both the crab crossing scheme and corresponding head-on scheme. In both schemes, a particle with small longitudinal action has a small diffusion index. For crab crossing scheme, more particles are driven by resonances and the largest diffusion index appears around the ellipse

$$\left(\frac{z}{2\sigma_z}\right)^2 + \left(\frac{p_z}{2\sigma_{p_z}}\right)^2 \approx 1 \qquad (34)$$

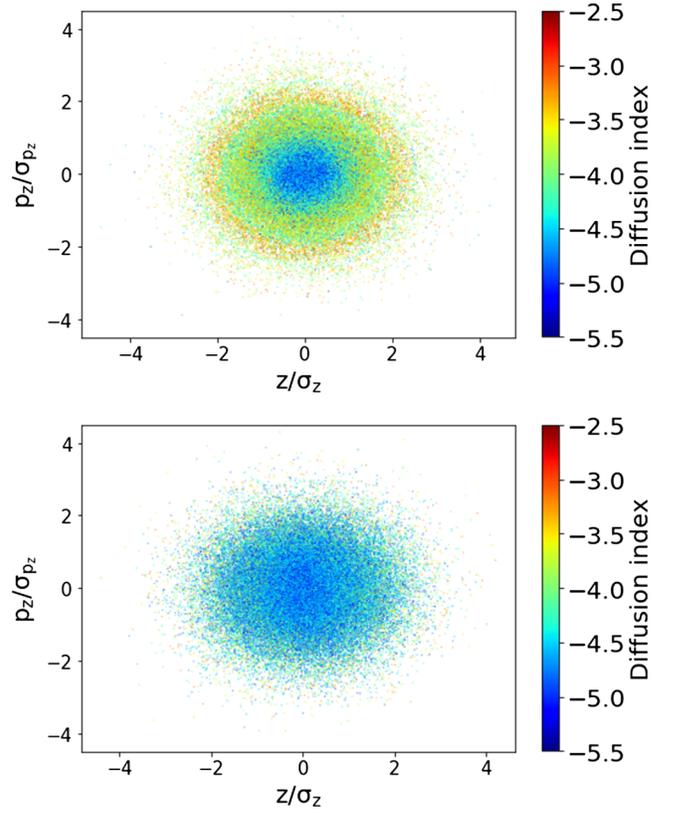

FIG. 16. Particle distribution in longitudinal phase space. Upper and lower pictures are for crab crossing and head-on collision scheme.

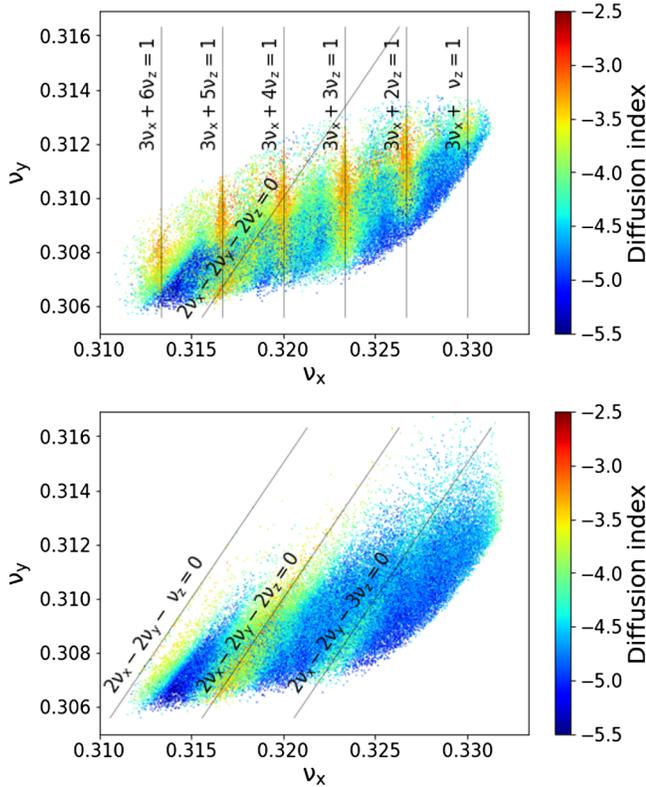

FIG. 15. The frequency map for nominal parameters in Table I, the top figure is for crab crossing scheme and the bottom figure is for head-on scheme. Particles with large diffusion index are labeled by red color while the small ones are in blue.

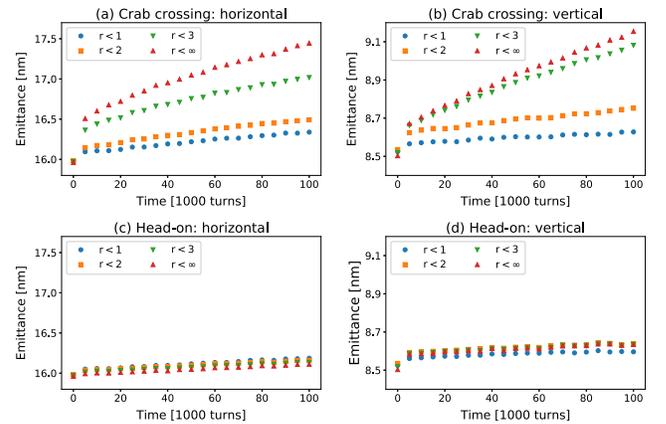

FIG. 17. Emittance evolution as function of turns, $r$ is defined by Eq. (35).

The ion macro-particles are then grouped according to their longitudinal action. Defining

$$r \equiv \frac{J_z}{\epsilon_z} = \frac{1}{2}\left[\left(\frac{z}{\sigma_z}\right)^2 + \left(\frac{p_z}{\sigma_{p_z}}\right)^2\right] \quad (35)$$

where $J_z$ is the longitudinal action and $\epsilon_z = \sigma_z \sigma_{p_z}$ is the average longitudinal action for all ions, the emittance evolution of each group is shown in Fig. 17. For crab crossing scheme, significant growth occurs when considering the particles $r > 2$.

In previous subsections, it has been demonstrated that a smaller beam-beam parameter or synchrotron tune of the proton beam will reduce the emittance growth and luminosity degradation. Figure 18 presents the frequency maps for smaller beam-beam parameter (70% of the nominal ones) and smaller longitudinal tune (from $\nu_z = 0.01$ to $\nu_z = 0.005$). Other parameters in Fig. 18 remain the same.

As shown in Fig. 7, and 11, the luminosity degradation in both situations are improved but their behavior in the frequency space is different. Compared with Fig. 15, The upper one in Fig. 18 occupies a smaller area in tune space so that less resonance lines are reached. The working area in the lower one in Fig. 18 is not reduced because the beam-beam tune shift is kept constant from Fig. 15. But resonance lines are narrower and weaker with smaller synchrotron tune. Especially, the second kind of resonances in Eq. (33) almost disappear from the footprint.

### D. Driving terms analysis

We use thin-strong electron model which is described in Sec. II to get the driving term strength of the synchrobetatron resonances. The truncated orders in Eq. (23) for horizontal and vertical plane are $M = 120$ and $N = 120$, respectively. In our numerical calculation, we first randomly choose 10,000 macroparticles from an exponential distribution,

$$\rho(J_x, \psi_x, J_y, \psi_y) = \frac{1}{4\pi^2 \epsilon_x \epsilon_y} \exp\left(-\frac{J_x}{\epsilon_x} - \frac{J_y}{\epsilon_y}\right) \quad (36)$$

where $J_{x,y}$ are the action variables, $\psi_{x,y}$ the phase factors, $\epsilon_{x,y}$ the proton beam emittances, and $\rho$ is the proton beam distribution. The driving terms $h_{3,0}$ and $h_{2,-2}$ are calculated from Eq. (25) for every macroparticle with different longitudinal coordinates. The average driving terms for these 10,000 macroparticles are shown in Fig. 19.

Figure 19 can be used to further explain the frequency maps Fig. 13–14 and the simulation results Fig. 3. In the crab crossing scheme, $|h_{3,0}|$ is far larger than $|h_{2,-2}|$. In other words, the first kind of resonances dominate the luminosity degradation. However, in the head on scheme, $h_{3,0}$ vanishes and then $h_{2,-2}$ becomes the dominated one. This explains that the luminosity degradation mitigates a lot when two beams collide head on. The $|h_{3,0}|$ reaches maximum when $|z| \approx 2\sigma_z$ which agrees with the Fig. 16, where the largest diffusion ellipse appears in the middle. It is also worthwhile to mention that the $h_{2,-2}$ looks quite different in the crab crossing and head on scheme. In the head on scheme, it is more likely a quadratic function of longitudinal $z$ and its strength is stronger. However in the crab crossing scheme, $h_{2,-2}$ disappears for large $|z|$.

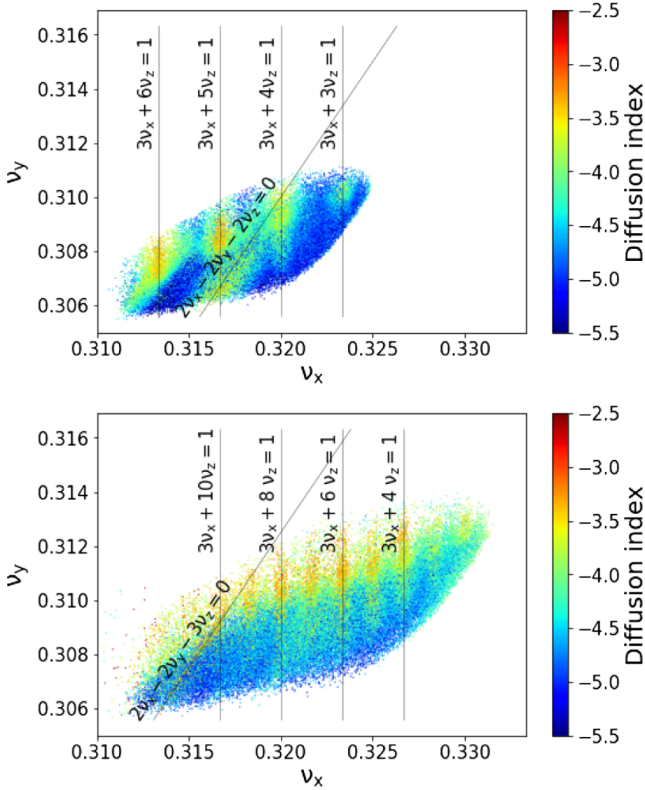

FIG. 18. Frequency maps tracked by SS simulation. The upper is for a smaller beam-beam parameter while the lower is for a smaller longitudinal tune.

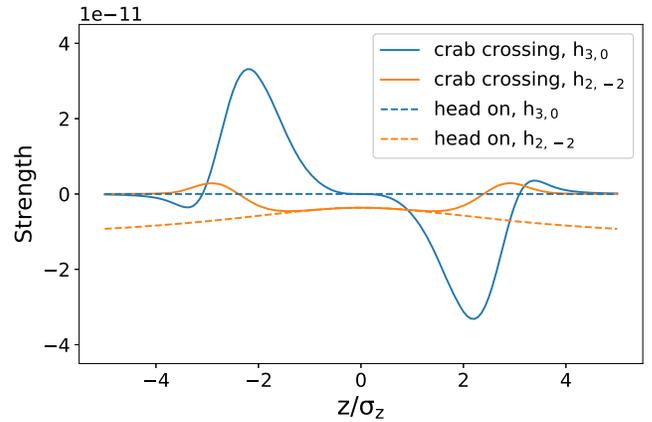

FIG. 19. The average driving terms versus longitudinal coordinates for crab crossing and head on collision scheme. The design parameters are used during calculation.

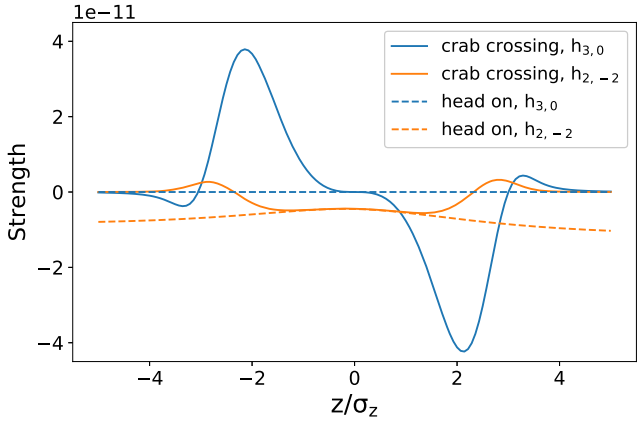

FIG. 20. The average driving terms versus longitudinal coordinates for crab crossing and head on collision scheme. The beam parameters are extracted from strong-strong simulation.

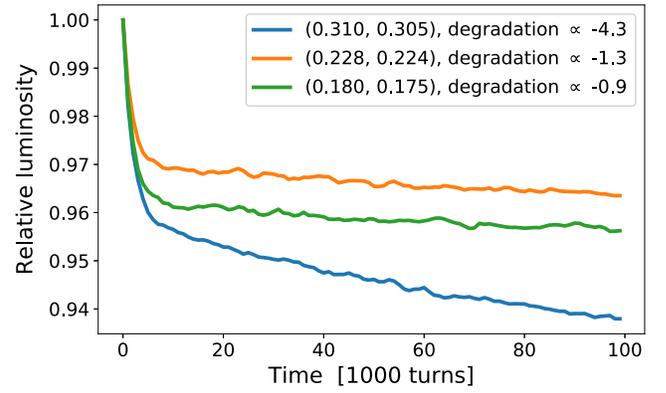

FIG. 21. SS simulation results for different tune choice of the proton beam. The longitudinal tune is kept as $\nu_z = 0.01$.

In Fig. 19, the $h_{3,0}$ has odd symmetry while the $h_{2,-2}$ has even symmetry with respect to $z = 0$. However the symmetry relations do not hold when the thin-strong electron slice is not at $z = 0$. Figure 20 shows similar calculation but the beam sizes and the Twiss parameters are extracted from the strong-strong simulation. If we consider multiple slices in our model, the asymmetry should be more obvious.

## IV. POSSIBLE MITIGATION METHODS

According to scaling studies in Sec. III B, the mitigation methods may include lowering the beam-beam parameter, the synchrotron tune, the bunch length and the crab cavity frequency of the ion beam. However, these approaches may not be ideal or limited by other aspect of constrains. For example, lowering the beam-beam parameter will impact the luminosity; the bunch length and synchrotron tune are limited by the rf voltage and longitudinal stability conditions, and the frequency of the crab cavity is limited by capability of the cavity manufacturing.

In this section, we will study three feasible methods to mitigate luminosity degradation while leaving luminosity less affected.

### A. Tune optimization

The most convenient method of mitigation is to choose the unperturbed working point below the diagonal resonance line $\nu_x = \nu_y$ (fractional tune) to avoid betatron and synchrobetatron resonances. Since the longitudinal tune $\nu_z$ is much smaller than the transverse ones, the second kind synchrobetatron resonances in Eq. (33) are unavoidable. However, the first kind of resonances could be weaker for higher order $|m| + |n|$. The degradation rate would be smaller if we chose the unperturbed working point $m\nu_x \lesssim 1, m > 3$.

Figure 21 compares the simulation results for different tunes near 1/3, 1/4 and 1/5 resonance respectively. The luminosity degradation is mitigated when the order of first kind resonance increases.

The proper choice of working points should consider all nonlinear components in the ring. This study indicates that the crab-cavity caused nonlinear resonance is also sensitive to the working point. It should be considered when a systematic optimization of the working point is performed.

A smaller longitudinal tune is also beneficial to the degradation as shown in Fig. 11. However, a smaller $\nu_z$ means a larger bunch length $\sigma_z$, which increases the finite bunch length effect, and possible instabilities.

### B. Higher harmonic crab cavity

A higher harmonic crab cavity can be used to reduce the tilting effect of proton beam. The crabbing offset in Eq. (18) turns into

$$f(z) = -\tan\theta_c \left[\frac{1+\alpha}{k_c}\sin(k_c z) - \frac{\alpha}{mk_c}\sin(mk_c z) - z\right] \quad (37)$$

here, $m$ is the harmonic number, and $\alpha$ is the relative strength of the harmonic cavity. More harmonics can be used to flatten the proton beam around IP in principle. However, more cavities not only make the crab cavity system complicated, but also poses great technical challenges in the low level rf control and fabrications of these cavities.

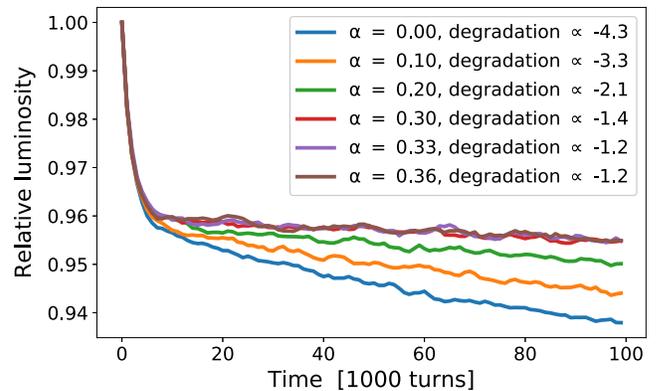

FIG. 22. SS simulation results in the presence of second-order harmonic crab cavity. $\alpha$ is the relative strength of the second-order harmonic crab cavity.

As a demonstration, Fig. 22 shows the luminosity evolution in the presence of a second-order harmonic crab cavity. The degradation rate reaches the minimum when $\alpha = 0.33$ as the coefficient of $z^3$ in the Taylor series of Eq. (37) vanishes.

### C. Dual interaction points

In the situation of two interaction points (IPs) in the EIC rings, we will demonstrate that the first kind synchrobetatron resonances in Eq. (33) can be cancelled or reduced by proper transformation between two IPs.

From Eq. (15) and (21), we have

$$\mathcal{M}_{IP} = \exp\left(-:a_{20}x^2 + a_{20}y^2 + a_{30}x^3 + \cdots :\right)$$

We omit $a_{00}$ and $a_{10}x$ here. We rewrite it into the following term

$$\mathcal{M}_{IP} = \exp\left(-:U_{odd} + U_{even}:\right) \tag{38}$$

where

$$U_{odd} = \sum_{m=0}^{\infty}\sum_{n=0}^{\infty} a_{mn} x^{2m+3} y^{2n}$$

$$U_{even} = \sum_{m=0}^{\infty}\sum_{n=0}^{\infty} a_{mn} x^{2m+2} y^{2n}$$

The term $U_{odd}$ can be eliminated by a $-\mathcal{I}_x$ transformation, which turns $(x, p_x, y, p_y)$ into

$$\begin{aligned}(-\mathcal{I}_x)x &= -x, & (-\mathcal{I}_x)p_x &= -p_x, \\ (-\mathcal{I}_x)y &= \pm y, & (-\mathcal{I}_x)p_y &= \pm p_y\end{aligned} \tag{39}$$

The map for two IPs combined by a $-\mathcal{I}_x$ map is

$$\begin{aligned}\mathcal{M}_{IP}\cdot(-\mathcal{I}_x)\cdot\mathcal{M}_{IP} &= (-\mathcal{I}_x)e^{:U_{odd}-U_{even}:}e^{-:U_{odd}+U_{even}:} \\ &= (-\mathcal{I}_x)e^{-:2U_{even}:}\end{aligned} \tag{40}$$

The last step holds because $:U_{odd}:$ and $:U_{even}:$ commute, i.e.,

$$[U_{odd}, U_{even}] = 0$$

Therefore, the synchrobetatron resonances excited by $U_{odd}$ are canceled by another IP and the first driving terms in Eq. (33) are eliminated.

A $-\mathcal{I}_x$ transformation can be made by

$$\Delta\psi_x = (2m+1)\pi, \qquad \Delta\psi_y = n\pi \tag{41}$$

where $m$, $n$ are integers, $\Delta\psi_{x,y}$ are the phase advance between two IPs in the horizontal and vertical plane.

Figure 23 demonstrates the cancellation effect for two IPs. The tracking is done by weak-strong code. In the simulation, the emittances of both beams are doubled to keep the luminosity and beam-beam parameters the same by summing the contribution of 2 IPs. The fractional tunes of the proton ring are kept at (0.310,0.305). From the above analysis, the horizontal betatron phase advance between the two IPs should be $\Delta\psi_x = \pi$. The case of $\Delta\psi_x = 2\pi$ is also calculated as a comparison. The vertical betatron phase advance is $\Delta\psi_y = \pi$. It is understandable that the degradation rate is much smaller in weak-strong simulation because there is less numerical noise in the model. We can see from Fig. 23 that the degradation rate for $\Delta\psi_x = \pi$ is negligible compared with the single IP or $\Delta\psi_x = 2\pi$. It proves that the $-\mathcal{I}_x$ map is an effective method to decrease the luminosity degradation rate. As another point of view, the result also supports that the degradation is caused by the synchrobetatron resonance.

Here we only focus on how to reduce the synchrobetatron resonance with two IPs. However, if we take the nonlinear chromaticity and cancellation into consideration, an intrinsic compensation requires a phase advance of odd times $\pi/2$ between the IPs [19]. It is also worthwhile to note that a lattice with two-fold symmetry may also suppress the degradation rate, and according to our preliminary simulation, has a similar result as the $\pi$ phase advance. Nevertheless, the dynamic beta effect of the symmetrical lattice, especially on the electron beam, is different from those of other examples in this section and needs further detailed studies. The real machine needs comprehensive consideration.

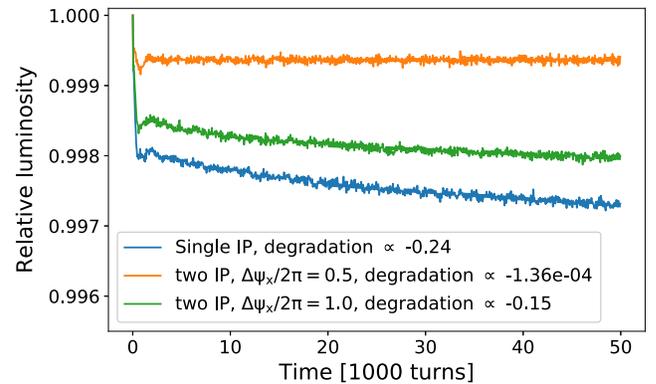

FIG. 23. Weak-strong simulation results for two IPs. The transverse fractional tunes of the proton ring are (0.310,0.305) in all three cases. Here the $\Delta\psi_x$ is the horizontal betatron phase advance between the IPs.

## V. CONCLUSION

We studied the colliding beam dynamics with a finite bunch length and a large crossing angle in the crab crossing scheme. A thin-strong model was applied to derive the beam-beam Hamiltonian in the presence of the crab cavities. When the beam-beam collision is modulated by the longitudinal dynamics, a single betatron resonance then splits into a spectrum of synchrobetatron resonances. The resonance driving terms of the synchrobetatron resonance are calculated numerically.

The strong-strong simulation revealed that the synchrobetatron resonance would cause ion beam's quality degradation and luminosity degradation. With the help of the frequency map and the driving term analysis, the dominated resonances were identified for an EIC like parameter set.

This paper also revealed that the resonances and degradation could be reduced or mitigated by modifying working points, reducing the synchrotron frequency, and adding harmonic cavities for the crab cavity. These studies provided useful inputs to optimize the parameter of a future electron-ion collider.

## ACKNOWLEDGMENTS

The authors would like to thank M. Blaskiewicz, W. Fischer, C. Montag, V. Ptitsyn and F. Willike for insightful discussions. This research used resources of the National Energy Research Scientific Computing Center (NERSC), a U.S. Department of Energy Office of Science User Facility operated under Contract No. DE-AC02-05CH11231. This work was supported by Department of Energy under award No. DE-SC0018973, and Contract No. DE-AC02-98CH10886.

## APPENDIX: EXPANSION OF BEAM-BEAM POTENTIAL

The first order differential of beam-beam potential $U(x,y)$ for a bi-Gaussian beam was derived by Bassetti and Erskine [20]

$$U_y + iU_x = -\frac{Q_1Q_2Nr_0}{\gamma_0}\sqrt{\frac{2\pi}{\sigma_x^2 - \sigma_y^2}}\left[w(z_2)\right.$$
$$\left. - w(z_1)\exp\left(-\frac{x^2}{2\sigma_x^2} - \frac{y^2}{2\sigma_y^2}\right)\right] \quad (A1)$$

where

$$U_x \equiv \frac{\partial U(x,y)}{\partial x}, \qquad U_y \equiv \frac{\partial U(x,y)}{\partial y}$$

$$z_1 = \frac{\frac{\sigma_y}{\sigma_x}x + i\frac{\sigma_x}{\sigma_y}y}{\sqrt{2(\sigma_x^2 - \sigma_y^2)}}, \qquad z_2 = \frac{x + iy}{\sqrt{2(\sigma_x^2 - \sigma_y^2)}}$$

In Eq. (A1), $w(z)$ is the Faddeeva function defined as

$$w(z) \equiv \exp(-z^2)\left(1 + \frac{2i}{\sqrt{\pi}}\int_0^z dt e^{t^2}\right) \quad (A2)$$

Its derivative is given by

$$w'(z) = \frac{2i}{\sqrt{\pi}} - 2zw(z) \quad (A3)$$

Taking the partial derivative on the both sides of Eq. (A1), we can get

$$U_{xy} = U_{yx} = \frac{-xU_y + yU_x}{\sigma_x^2 - \sigma_y^2} \quad (A4)$$

$$U_{xx} = -\frac{xU_x + yU_y}{\sigma_x^2 - \sigma_y^2} - \frac{2Q_1Q_2Nr_0}{\gamma_0(\sigma_x^2 - \sigma_y^2)}$$
$$\times \left[1 - \frac{\sigma_y}{\sigma_x}\exp\left(-\frac{x^2}{2\sigma_x^2} - \frac{y^2}{2\sigma_y^2}\right)\right] \quad (A5)$$

$$U_{yy} = \frac{xU_x + yU_y}{\sigma_x^2 - \sigma_y^2} + \frac{2Q_1Q_2Nr_0}{\gamma_0(\sigma_x^2 - \sigma_y^2)}$$
$$\times \left[1 - \frac{\sigma_x}{\sigma_y}\exp\left(-\frac{x^2}{2\sigma_x^2} - \frac{y^2}{2\sigma_y^2}\right)\right] \quad (A6)$$

Substituting $U_{x,y,xx,yy}$ with their Taylor representation

$$U_x = \sum_{m,n=0}^{\infty}(m+1)a_{m+1,n}x^m y^n$$

$$U_y = \sum_{m,n=0}^{\infty}(n+1)a_{m,n+1}x^m y^n$$

$$U_{xx} = \sum_{m,n=0}^{\infty}(m+2)(m+1)a_{m+2,n}x^m y^n$$

$$U_{yy} = \sum_{m,n=0}^{\infty}(n+2)(n+1)a_{m,n+2}x^m y^n$$

and $x$ with $x + f$ into Eq. (A5) and Eq. (A6), we can get the recursion expression about $a_{mn}$

$$a_{m+2,n} = -\frac{(m+1)a_{m+1,n}f + (m+n)a_{mn}}{(m+2)(m+1)(\sigma_x^2 - \sigma_y^2)}$$
$$- \frac{2Q_1Q_2Nr_0}{\gamma_0(\sigma_x^2 - \sigma_y^2)}\frac{t_{mn}^{(1)}}{(m+2)(m+1)} \quad (A7)$$

$$a_{m,n+2} = \frac{(m+1)a_{m+1,n}f + (m+n)a_{mn}}{(n+2)(n+1)(\sigma_x^2 - \sigma_y^2)}$$
$$+ \frac{2Q_1Q_2Nr_0}{\gamma_0(\sigma_x^2 - \sigma_y^2)}\frac{t_{mn}^{(2)}}{(n+2)(n+1)} \quad (A8)$$

where $t_{mn}^{(1)}$ and $t_{mn}^{(2)}$ satisfy

$$1 - \frac{\sigma_y}{\sigma_x}\exp\left[-\frac{(x+f)^2}{2\sigma_x^2} - \frac{y^2}{2\sigma_y^2}\right] = \sum_{m,n=0}^{\infty} t_{mn}^{(1)} x^m y^n$$

$$1 - \frac{\sigma_x}{\sigma_y}\exp\left[-\frac{(x+f)^2}{2\sigma_x^2} - \frac{y^2}{2\sigma_y^2}\right] = \sum_{m,n=0}^{\infty} t_{mn}^{(2)} x^m y^n \quad \text{(A9)}$$

and they can be calculated to very high orders by using the truncated power series algebra (TPSA) technique.

The initial condition for the recursion expression is

$$a_{01} + i a_{10} = -\frac{Q_1 Q_2 N r_0}{\gamma_0}\sqrt{\frac{2\pi}{\sigma_x^2 - \sigma_y^2}}$$
$$\times \left[w\left(\frac{f}{\sqrt{2(\sigma_x^2 - \sigma_y^2)}}\right)\right.$$
$$\left. - w\left(\frac{f\sigma_y/\sigma_x}{\sqrt{2(\sigma_x^2 - \sigma_y^2)}}\right)\exp\left(-\frac{f^2}{2\sigma_x^2}\right)\right] \quad \text{(A10)}$$

We can get $a_{mn}$ by using Eq. (A7)–(A10)). However, $a_{mn}$ becomes very large as $m$ and $n$ increase. It is necessary to use high precision floating-point in our code.